\documentstyle[amssymb,aps,epsfig,multicol]{revtex}
\begin{document}
\title{Fluctuation Magnetoconductance in MgB$_{2}$}
\author{W. N. Kang, Kijoon H. P. Kim, Hyeong-Jin Kim, Eun-Mi Choi,
        Min-Seok Park, Mun-Seog Kim, Zhonglian Du, Chang Uk Jung, Kyung Hee Kim, and
        Sung-Ik Lee\footnote{silee@postech.ac.kr}}
\address{National Creative Research Initiative Center for Superconductivity and Department of Physics,
Pohang University of Science and Technology, Pohang 790-784,
Republic of Korea}
\author{Mi-Ock Mun}
\address{Institute of Physics and Applied Physics and Department of Physics,
         Yonsei University, Seoul 120-749, Republic of Korea }
\date{\today }

\maketitle
\begin{abstract}
We report fluctuation magnetoconductance in a MgB$_{2}$
superconductor prepared under a high pressure of 3 GPa. This
excess magnetoconductance, $\Delta\sigma(H)$, follows the
three-dimensional scaling function for the critical fluctuations
proposed by Ullah and Dorsey. This feature is inferred to
originate from the isotropic nature of the MgB$_2$ compound.
\end{abstract}
\pacs{PACS number: 74.25.Fy, 74.70.Ad, 74.25.Jb, 74.70.-b}

\begin{multicols}{2}
The recent discovery of superconductivity at about 40 K in
MgB$_{2}$~\cite{akimitsu1} brought about an avalanche of research.
While the type of carrier was found to be
hole-like~\cite{wnkang7}, other properties have been reminiscent
of conventional BCS superconductivity, and are significantly
different from those found in the cuprate superconductors. These
properties include a notable shift in the $T_{c}$ due to the boron
isotope,~\cite{budko7,Kortus7}, a stable metallic behavior with a
higher carrier density~\cite{wnkang7}, and prominent bulk
pinning\cite{mgb2mskim7}.

It is well-known that high-$T_c$ superconductors have a
two-dimensional layered structure and a very short out-of-plane
coherence length, which is quite different from conventional
superconductors. Thus, a strong superconducting fluctuation effect
around $T_c$ is evident for thermodynamic and transport
quantities. In the MgB$_2$ system, various superconducting
properties, including the electrical conductivity, are believed to
be isotropic even though the boron planes are thought to act like
the CuO$_{2}$ planes in the cuprate
superoconductors~\cite{hirsch7}. Thus, it could be very
interesting to investigate the conductivity fluctuation effect in
this compound.

In this paper, we present the magnetoconductance fluctuation in a
MgB$_{2}$ polycrystal. The scaling functions for the critical
fluctuations were employed to analyze the excess conductance of
our sample. We found that the conductivity in the critical region
scales with the three dimensional scaling functions.

A 12-mm cubic multi-anvil press was used for the high-pressure
sintering. The pellet, which was made of a commercially available
powder of MgB$_{2}$\cite{memo4}, was sintered at about $900^{\circ
}$C after being subjected to pressures of up to $3$ GPa. The
zero-field resistivity transition temperature, $T_{c}^{R}(0)$, is
38.3 K. A transmission electron microscope (TEM)
study~\cite{sung7} showed that all grains were compactly connected
and that no discernable empty space or impurities at the
boundaries. The weak link behavior is expected to be less than
that observed in to the high-T$_c$ superconductors. The details of
the sample preparation and the properties of sample can be found
elsewhere.\cite{cujung1}

The voltage noise was successfully reduced to a lower level by
preparing a very thin and optically clean specimen polished from a
strong, dense bulk pellet. The dimensions of the specimen were
about $2.4\times 4.0\times 0.1$ mm$^{3}$. The standard
photolithography technique was used to make electrical leads on
the specimen. To obtain good ohmic contacts between the Au pads
and the sample, we cleaned the surface using an ion beam before
depositing Au-contact pads. The magnetic field was applied
perpendicular to the sample surface and the current direction. The
resistivity measurements were carried out in a cryostat with a
superconducting magnet by using the standard 4-probe technique. A
dc current source (model Keithley 220), two channel nanovoltmeters
(model HP 34420A), and dc preamplifier (model EM electronics N11)
with $J = 40-50$ A/cm$^{2}$ were used.

\begin{figure}
\centering \epsfig{file=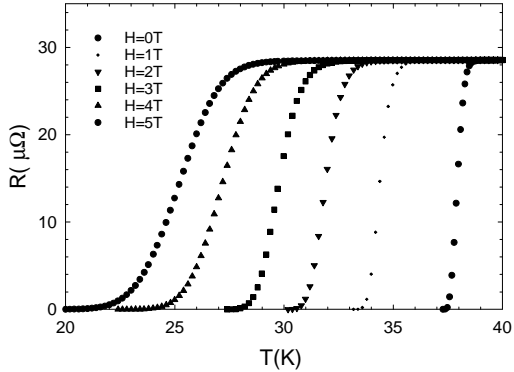,width=7cm}
\caption{Magnetic-field dependence of resistitance of MgB$_{2}$
         sintered under 3 GPa. }
\label{fig1}
\end{figure}

Figure~\ref{fig1} shows the temperature dependence of the
resistance for applied magnetic fields of 0 T $\leq H \leq$ 5 T.
The gradual change in the resistance near the transition
temperature is similar to the case of typical metallic
superconductors. The resistance just above $T_{c}$ did not change
with the applied field. It is clearly seen that the resistivity
curves shift toward lower temperature as the field increases. The
rate of decrease of $T_c(H)$ with respect to the field is
significantly larger than that of the cuprates. The transition
temperature $T_c(H)$ was estimated from the 90\% drop of the
normal-state resistivity, and the slope $(dH_{c2}/dT)|_{T_{c}}$
was found to be about -0.54 T/K. This value of the slope is much
smaller than the $dH_{c2}/dT$ of $\sim-2$ T/K observed in
high-temperature superconductors such as YBa$_2$Cu$_{3}$O$_7$ and
Bi$_2$Sr$_2$CaCu$_2$O$_8$. Using this slope, the estimated
$H_{c2}(0)$ was found to be 13.9 T through the relationship
$H_{c2}(0)\simeq0.7 T_c (dH_{c2}/dT)_{T_c}$,~\cite{werthamer1}
this value is about ten times smaller than the value for cuprate
superconductors. The coherence length $\xi(0)=[\phi_0/2\pi
H_{c2}(0)]^{1/2}$ was calculated to be 48.6 \AA, where $\phi_0$ is
the flux quantum.

In the strong-field region, since the quasi-particles are mainly
confined to their lowest Landau level,~\cite{Leep1} an effective
reduction of the dimensionality occurs, hence the
fluctuation-induced conductance exhibits a critical behavior.
Critical phenomena and scaling properties of the thermodynamic
quantities near the transition temperature in a mixed state have
been studied by a number of
groups.~\cite{bray1,thouless1,ruggeri1} Ullah and
Dorsey\cite{ullah1} proposed scaling functions for various
thermodynamic and transport quantities within the Hartree
approximation. The quartic term $|\psi|^4$ in the Ginzburg-Landau
free energy was included which was in contrast to the free
fluctuation theory where the quartic term is neglected. The
scaling functions for the excess conductance are given by
\begin{equation}
\Delta \sigma (H)_{2D}=\left[ {\frac{T}{H}}\right] ^{1/2}{F_{2D}}\left[
{A\frac{{T-T_{c}(H)}}{{(TH)^{1/2}}}}\right] ~~~\mbox{for 2D,}
\end{equation}
\begin{equation}
\Delta \sigma (H)_{3D}=\left[ {\frac{T^{2}}{H}}\right] ^{1/3}{F_{3D}}
\left[B{\frac{{T-T_{c}(H)}}{{(TH)^{2/3}}}}\right] ~~~\mbox{for 3D,}
\end{equation}
where $F_{2D}$ and $F_{3D}$ are unspecified scaling functions, and
$A$ and $B$ are field and temperature-independent coefficients. In
these equations, the Aslamazov-Larkin(AL) term~\cite{Aslamazov1}
of the fluctuation conductance was taken into account.

\begin{figure}[tbp]
 \centering \epsfig{file=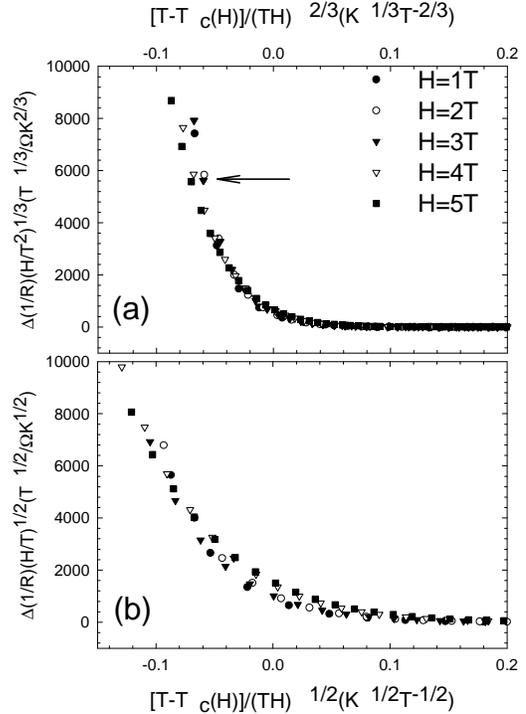,width=7cm}
\caption{(a) 3D scaling fit of the magnetoconductance fluctuation.
         (b) 2D scaling fit of the magnetoconductance fluctuation.}
\label{fig2}
\end{figure}

To obtain the excess conductance $\Delta\sigma (H)$, we
substracted the extrapolated normal conductance from the total
conductance by using a normal fit at temperatures far above
$T_{c}(H)$ for each magnetic field. The correct determination of
the $T_{c}(H)$ is critical for obtaining the best scaling
behavior. For $T_{c}(H)$, we used the refined temperature around
the 10\% resistive superconducting transition point for each
field. The excess conductance scaled excellently with the 3D form.
Figure~\ref{fig2}(a) shows $\Delta(1/R)/(T^2/H)^{1/3}$ versus the
scaling parameter $(T-T_c(H))/(TH)^{2/3}$. All the data for the
various fields collapse into a single curve while the 2D scaling
plot gives a poor result, as shown in Fig.~\ref{fig2}(b), for
whatever value of $T_c(H)$. The values of $T_c(H)$ for the best
fits were 34.9, 32.8, 31.0, 28.8, and 27.2 K for H = 1, 2, 3, 4,
and 5 T, respectively, which slightly deviated from linearity in H
but agreed with the generalized prediction by Maki.~\cite{Maki3}

It is quite surprising to find 3D scaling behavior in a
polycrystalline sample. A compactly connected morphology with no
discernable empty space was observed using a TEM, thus, the weakly
linked grain boundary effect is negligible. The high degree of
iosotropy is another feature for this 3D scaling behavior. The
spreadout of the scaled conductance at low temperatures, indicated
by the arrow in Fig.~\ref{fig2}(a), is not unusual for the scaling
behavior in the fluctuation conductivity. Several results obtained
from high-$T_c$ superconductors~\cite{pradhan1}, as well as
$R$Ni$_{2}$B$_{2}$C materials~\cite{momun1} have commonly given
the same broadening due to the vortex motion at the tail of the
transition line \cite{pradhan1}.

In summary, we observed magnetoconductance fluctuations in
MgB$_{2}$. With critical fluctuations  includede, the scaling
behavior of MgB$_{2}$ was better explained by the 3D theory rather
than by the 2D theory. Even though the grains in the sample are
randomly oriented, the excess conductance follows the 3D scaling
function for critical fluctuations. This feature may indicate that
MgB$_2$ is nearly isotropic. However, more rigorous studies of the
suggested 3D nature of MgB$_{2}$ require at least a grain-aligned
sample such as a $c$-axis oriented film or a single crystal.

This work is supported by the Ministry of Science and Technology of Korea
through the Creative Research Initiative Program.

\end{multicols}
\end{document}